\magnification=\magstep1
\baselineskip=18pt
\hfuzz=6pt

$ $

\rightline{Los Alamos 1992}
\rightline{LA-UR-92-996}

\vfill

\centerline{\bf Any non-affine one-to-one binary gate
suffices for computation}

\vskip 1cm

\centerline{Seth Lloyd}

\bigskip
\centerline{T-13, Center for Nonlinear Systems} 

\centerline{ Los Alamos National Laboratory,
Los Alamos, New Mexico 87545}

\vskip 1cm 

\noindent{\it Abstract:} Any non-affine one-to-one binary gate
can be wired together with suitable inputs to give $AND$, $OR$, $NOT$
and fan-out gates, and so suffices to construct a general-purpose
computer.

\vfill\eject

\noindent{\it Introduction} 

Since the discovery [1] that the process
of computation can in principle be carried out reversibly, without
dissipation, a number of designs for reversible computers have
been proposed [2-5].  At the heart of many such designs lies a reversible logic
gate, that is capable of performing basic logical operations in a manner
that discards no information: the output of such a gate is a one-to-one
function of its input.   A number of physical systems have
been proposed to realize such logic gates using, for example, a classical
hard sphere gas in a periodic potential [4], or nonlinear optics [6-7].

For the moment, such proposals are far from realizing a working, non-dissipative
gate that exhibits sufficient stability and noise resistance to be wired
together in a general-purpose computer [8-9].   Even if such computers could
be made to work, they would have a hard time initially, competing with
the present day's remarkably speedy and efficient semiconductor based
machines.    But semiconductors have limits on maximum speed and
minimum dissipation.   If these limits are to be surpassed, some day, new
technologies must be made available.

The present paper shows that any physical process that can give a 
non-affine one-to-one binary logic gate can serve as a basis
for constructing a computer.   In a companion paper [10], it is shown
that if linear operations are supplemented by any nonlinear gate
at all, the resulting set of operations suffices for computation. 
Whether these results can bacilitate actually building working computers is an 
open question.   However, no nonlinear effect can be ruled out 
{\it a priori} as a basis for computation.

\bigskip\noindent{\it One-to-one logic gates}  

The original one-to-one
logic gate to be proved to suffice for computation is
the Fredkin gate [4].   This gate has three binary inputs, $x,y,z$ and
three binary outputs $x',y',z'$.  The first input passes through
unchanged: $x'=x$.   If the first input is zero, then the scond
and third inputs are passed through unchanged:
$x=0 \rightarrow y'=y, z'=z$.   If the first input is one,
the second and third inputs are interchanged: $x=1 \rightarrow
y'=z, z'=y$.  The Fredkin gate is clearly one-to-one.   To suffice for 
computation, several copies of the gate must be able to
be wired together, with some inputs fixed to certain values,
to function as $AND$, $OR$, $NOT$, and fan-out gates (a fan-out
gate is one that outputs two copies of its input).
$AND$, $OR$, $NOT$ and fan-out gates form a basis for computation, in 
the sense that copies of these gates can be wired together to give
any desired logic circuit.

\smallskip\noindent(1) Fan-out: If the second and third inputs are fixed 
to zero and one respectively, then the first and second outputs are copies
of the first input.    That is, $y=0$, $z=1 \rightarrow
x'=x, y'=x$.   So the Fredkin gate can be used to make copies of bits.

\smallskip\noindent(2) $NOT$: With the same inputs as for constructing
the fan-out gate, the third output is
equal to the negation of the first input: $z'= NOT~x$.

\smallskip\noindent(3) $AND$: If the third input is fixed to zero, then
the third output is one if and only if the first and second inputs are
one, and zero otherwise: $z=0 \rightarrow z' = x~AND~y$.

\smallskip\noindent(4) $OR$: If the third input is fixed to one, then the 
second output is zero only if the first and second inputs are zero,
and one otherwise: $z=1 \rightarrow y' = x ~OR ~ y$.

A set of gates is a basis for computation if gates form the set can
be wired together to realize any logic circuit.   Since $AND$, $OR$,
$NOT$ and fan-out gates can be wired together to realize any
logic circuit, the Fredkin gate alone can be wired together with
some inputs fixed to create an arbitrary circuit, and so forms
a basis for computation.

\bigskip\noindent{\it Any non-affine one-to-one binary gate
suffices for computation}

The result is proved in three steps.   First, it is shown that
any non-affine binary gate, together with $NOT$ and fan-out, gives
a basis for computation.   Second, any
injective non-affine binary gate (an injective gate is one for which
no two sets of values for the inputs gives the same set of values
for the outputs) is shown to give a basis for computation
when combined with $NOT$ alone.   Third and finally, any one-to-one
non-affine binary gate is proved to provide a basis for
computation on its own. 

Consider an arbitrary $n$-input, $m$-output, non-affine binary gate,
with inputs $x_1, \ldots, x_n$ and outputs $x'_1, \ldots, x'_m$.
An affine function of $x_1, \ldots, x_n$ can be written
$f(x_1, \ldots, x_n) = a_0 + a_1 x_1 + \ldots + a_nx_n$, where
addition and multiplication are devined modulo 2.  An $n$-input
$m$-output gate is non-affine if at least one of its outputs is
not an affine function of its inputs.

\bigskip\noindent (1) Any $n$-input, $m$-output non-affine
binary gate can be combined with $NOT$ and fan-out to give a
basis for computation.   

\smallskip\noindent{\it Proof:} By induction on the number of
inputs.  Inductive hypothesis: Assume that any $k-1$ input non-affine
binary gate can be combined with $NOT$ to give $AND$ and $OR$.
This hypothesis is true by inspection for 2-input gates: such gates must
have at least one output that is a non-affine function of its inputs,
and any 2-input, 1-output gate can be combined with $NOT$ to give
$AND$ and $OR$.   Now consider a $k$-input non-affine gate:
one of the gate's outputs $x'$ is a non-affine function of the inputs.
There are two cases to consider:

\item{i.} The output, $x'$, is a non-affine function of $k-1$ inputs
for some value of the $k$th input.   In this case, the gate suffices 
to construct $AND$ and $OR$ gates by the inductive hypothesis.

\item{ii.} The output, $x'$ is an affine function of $k-1$ inputs
for any value of the $k$th input:
$$\eqalign{ x' &= a_0 + a_1 x_1 + \ldots + a_{k-1} x_{k-1} ~ {\rm for}
~ x_k=0\cr
 x' &= b_0 + b_1 x_1 + \ldots + b_{k-1} x_{k-1} ~ {\rm for}
~ x_k=1.\cr}$$
The only way that $x'$ can be a non-affine function of the inputs is
for at least one of the $a_i \neq b_i$ for $1\leq i \leq k-1$.   But
in this case, $x'$ is a nonlinear function of $x_i$ and
$x_k$ alone, for any values of the other inputs.    The gate can
then be used as a 2-input, non-affine binary gate with inputs $x_i,x_k$
and output $x'$, which can be combined with $NOT$ to give $AND$
and $OR$.

\smallskip\noindent So 
by induction, any $n$-input non-affine binary gate can be combined
with $NOT$ to give $AND$ and $OR$.   Since $NOT$, $AND$, $OR$ and fan-out give
a basis for computation, any non-affine gate together with $NOT$
and fan-out gives a basis for computation.

\bigskip\noindent(2) Any non-affine injective binary gate together
with $NOT$ gives a basis for computation.

The idea of this proof is simple: it is shown that any non-affine
injective gate can be combined with $NOT$ to give a fan-out gate.   Since
any non-affine gate combined with $NOT$ and fan-out suffices
for computation, as in (1) above, any non-affine injective
gate together with $NOT$ gives a basis for computation.

Fan-out: By the proof of (1), any injective non-affine binary gate
can be combined with $NOT$ to create a logic circuit in which all inputs
but two are fixed, and one of the outputs is the $AND$ of the two
inputs that are varied.   Since the gate is injective, the
resulting circuit is also injective, and one must be able to 
recreate the values of the two variable inputs from looking
at the values of the outputs when those inputs are
varied.   A 2-input injective gate that includes $AND$ as one of
its outputs must have at least three outputs that vary when
the two inputs vary.  One can prove by inspection that for
some value of one of the inputs of a 2-input, $m$-output injective
gate that has $AND$ as one of its outputs, varying the other input must
induce a correlated variation in at least two of the outputs.

But any gate that has at least two outputs vary with one of its
inputs can be combined with $NOT$ to make a fan-out.   So an injective
non-affine binary gate can be combined with $NOT$ to give
a fan-out, and hence a basis for computation.

\bigskip\noindent (3) Any one-to-one non-affine binary gate gives 
a basis for computation.

To prove this, one need only show that any one-to-one non-affine
binary gate can be used to realize a $NOT$ gate.  (2) above
then implies that such a gate provides a basis for computation.

To be one-to-one, an $n$-input non-affine binary gate must have
$n$ outputs, as well.    To realize a $NOT$ gate, one need simply
exhibit $a_1, \ldots, a_{n-1}$ such 
that fixing $n-1$ of the inputs to these values, and varying
the remaining input as $x$, causes some one
of the outputs to vary as $NOT ~x$.  In fact, the only one-to-one
binary gates, affine or non-affine, that do not suffice to realize
a $NOT$ gate are those for which each input is
passed through to some output unchanged.

Consider the input-output table for the gate, in which the inputs
are listed as binary numbers in ascending order.   If any
column in the output table has a 1 at position $r$ and a 
0 at position $2r$, then the gate can be used to realize
a $NOT$ gate.   But if the gate is one-to-one, each column in the output
part of the table must have an equal number of zeros and ones.
The only columns that have equal numbers of ones
and zeros, and in no place a 1 at position $r$ and a 0 at
position $2r$ are of the form,
$010101 \ldots$, $001100110011 \ldots$, $0000111100001111\ldots$,
etc.    That is, the only one-to-one binary gates that connot be
used to give $NOT$ gates are gates, each of whose outputs
is equal to some one of its inputs.    Since such gates are
trivially affine, any non-affine one-to-one binary gate realizes
$NOT$.   By (1) and (2) above, any $n$-input, $n$-output non-affine
one-to-one binary gate suffices to construct $NOT$, $AND$, $OR$ 
and fan-out gates, and so gives a basis for computation.

\bigskip\noindent{\it Conclusion}

The results here are restricted to binary inputs and outputs.
Devices with discrete inputs that can take on more than two values,
and devices with continuous inputs behave somewhat differently.
It is no longer the case, for example, that any
non-binary non-affine device gives a basis for computation on
its own.   The reason is simple: with inputs and outputs taking on more
than two values, devices with only one input and one output
can be non-affine.   Such devices have the wrong number of inputs and
outputs to give $AND$, $OR$, or fan-out gates.   One can show,
however, that any non-affine device, discrete or continuous,
can be combined with fan-out and suitable linear devices
to give a basis for computation [10].

A further obstacle in applying the results derived here to
physical systems is that the real, microscopic systems that might
be used to construct fast, efficient computers are invariably
subject to fluctuations and noise.   Although von Neumann's
multiplexing technique can be employed to make the sort of gates
discussed here reliable in the face of small amounts of
noise [11], the conditions under which arbitrary noisy non-affine
gates can be combined to construct arbitrary logic circuits
are not known.

\vfill\eject
\noindent{\it References:}

\vskip 1cm

\smallskip\noindent [1] C.H. Bennett, {\it IBM J. Res. Develop.} 
{\bf 17}, 525-532 (1973).

\smallskip\noindent [2] P. Benioff, {\it Phys. Rev. Lett.}
{\bf 48}, 1581-1585 (1982).

\smallskip\noindent [3] R.P. Feynman, {\it Opt. News.} {\bf 11},
11-20 (1985).

\smallskip\noindent [4] E. Fredkin and T. Toffoli, 
{\it Int. J. Theor. Phys.} {\bf 21}, 219-253 (1982).

\smallskip\noindent [5] D. Deutsch, {\it Proc. Roy. Soc. Lond. A} {\bf 400},
97-117 (1985).

\smallskip\noindent [6] G.J. Milburn {\it Phys. Rev. Lett.}
{\bf 62}, 2124-2127 (1989).

\smallskip\noindent [7] Y. Yamamoto, M. Kitagawa, and K. Igeta, in
{\it Third Asia/Pacific Physics Conference,} C.N. Ying, Y.W. Chan,
K. Young, A.F. Leung, eds., World Scientific, Singapore (1988).

\smallskip\noindent [8] R. Landauer, in {\it Nanostructure
Physics and Fabrication,} M.A. Reed, W.P. Kirk, eds.,
Academic Press, San Diego (1989).

\smallskip\noindent [9] R.W. Keyes, {\it Science} {\bf 230},
138-144 (1985).

\smallskip\noindent [10] S. Lloyd, {\it Phys. Lett. A}
{\bf 167}, 255-260 (1992).

\smallskip\noindent [11] J. von Neumann, {\it Probabilistic logics
and the synthesis of reliable organisms from unreliable components,}
lectures delivered at the California Institute of Technology
(1952).

\vfill\eject\end

on
\vfill\eject\end